# A Model of an E-Learning Web Site for Teaching and Evaluating Online.

Mohammed A. Amasha
College of Science and Arts, Computer Science
Department, Qassim University
Alrass City, Saudi Arabia –Domyat University, Egypt

Salem Alkhalaf
College of Science and Arts, Computer Science
Department, Qassim University
Alrass City, Saudi Arabia

*Abstract*—This research is endeavoring to design an e-learning web site on the internet having the course name as "Object Oriented Programming" (OOP) for the students of level four at Computer Science Department (CSD). This course is to be taught online (through web) and then a programme is to be designed to evaluate students' performance electronically while introducing a comparison between online teaching, e-evaluation and traditional methods of evaluation. The research seeks to lay out a futuristic perception that how the future online teaching and e-electronic evaluation should be the matter which highlights the importance of this research.

*Keywords—Key Words; e-learning; evaluation; designing a website; university performance*

I. INTRODUCTION

The twenty-first century has witnessed the transformation of a new world in which contemporary information technology prevails in global society; where the educated people have a free access to the information they need within a short time with least efforts-the matter that results in enhancing the competence level of creativity and production [4].

The issue of educational development has topped the agenda of several countries on their conviction that the progress and promotion of their people begin with development of education. Hence, there are certain countries which have begun to amend their educational systems in their staunch faith that education is the only way to rise up to the developments and challenges characterizing the present era. These countries have sought to introduce modern technology to their educational system representing computer technologies in their bid to benefit from the potentials of computer as an educational aid and technological coordinator [20].

Still, a plethora of programmers and programming languages that aim at best optimizing the web in the domains of e-mails, e-conferences, and written/ voice-chat which have kept up with the emergence of the internet and its development. Thus, introducing the internet into education that has been given due care to help give students quick and easy access to knowledge [15].

The internet has remarkably and considerably contributed to the development of all aspects of the educational process. It doesn't only have a great bearing on educational curricula and methodology but it also extends to administrative domains. In addition, it helps to give a boost to the modern educational types as open education, distance education, self-learning and individual learning [17]. The benefits that internet can render are countless as it may bring about a genuine and effective development in Arab Educational Systems [15].

Educational development starts after accomplishing integrated procedures shaped into the development of a system, the foremost of which is giving due care to amending, improving educational curricula and methodology. Universities should be pioneers in developmental process as they are considered the locomotives of advancement and modernization in a society. Their message is to steer the process of development in the different institutions scientific knowledge and modern technology.

Due to this inspiration the educational process began to develop in all its spheres. In this context, some educational centres and units were established aiming at promoting the instructional, educational and research processes in order to achieve the goals of the ongoing development in an age of informational and developmental change. One of these units is E-learning Unit which has been playing the greatest role of developmental process.

This unit aims at the developing of the educational curricula in all specializations of the different colleges of the university in its pursuit to enhance the role of education in the realization of outputs of educational process. E-education helps to use strategies of competence and feedback in recognizing students' individual needs and devising ways to develop them and it also contributes to provide an intellectual forum, which aims to enhance the educational content and give a boost to the educational curricula. [1].

The current research attempts to design an an e-learning web site on the internet as a course namely, "Object Oriented Programming" (OOP) for the students of 4th level at CSD. OOP is a programming paradigm that represents concepts as "objects" that have data fields (attributes that describe the object) and associated procedures known as methods. Objects, which are instances of classes, are used to interact with one another to design applications and computer programs (www.wikipadeia.com).

This course is to be taught electronically (via the web) and then a programme is to be designed to e-valuate students and draw a comparison between e-evaluation and the traditional methods of evaluation. The research seeks to put forward a futuristic perception to the way future e-evaluation should be, the matter that highlights the importance of the research.

## II. STATEMENT OF THE PROBLEM

The previous presentation demonstrates the importance of using E-learning in university education, the point that has been sustained by several studies carried out in this field, which stresses on the importance of disseminating the culture of education amongst the teachers and their assistants, developing university e- courses and making the students best optimize these courses.

These studies highlight the importance of transforming evaluation methods to cope with the development that's taking place. Thus, change of test patterns has become a necessity as it is one of the characteristics of this age and a prerequisite to keep abreast with the successive developments of today's world. In addition, evaluation is considered a substantial element in the educational curriculum and it's, therefore, one of the foundations of education development because any development in the aims, content and methodologies of a given educational curriculum that cannot be effected without reliance on evaluation results. It's a must to introduce new techniques in the bases and sources of evaluation via modern technology in order to promote and develop tools of evaluation.

There upon, employing that e-evaluation has become a pressing need as it guarantees the credibility to scores and it secures the grader to be unprejudiced. Besides, e-evaluation helps to solve the problems of the staff represented in giving tests and, grading them, assessing students' performance especially in the light of the appalling number of students overcrowding the colleges of the university at present. Hence, the importance of e-evaluation lies. The question can now be phrased as follows:

Can e-evaluation be effective and fair with students? Can we safely state that the time is now ripe to manage without the traditional (paper) evaluation and to resort to e-evaluation? Is there an effective vision to the future form of e-evaluation? And do students have a real interest in using E-evaluation? All these significant points need to be made clear. The current research attempts to make a comparison between e-evaluation and the traditional methods of evaluation and to survey the students' opinions in connection with e-evaluation & employing the students' opinions poll to form a proposed future vision of e-evaluation.

## III. THE RESEARCH PROBLEM CAN BE COUCHED IN THE FOLLOWING QUESTIONS:

- Do e-evaluation techniques result in an improvement on the attainment level of the sample students?
- Does the use of e-evaluation method takes less time than the conventional method?
- What do the students' think of the proposed programme using an e-evaluation technique?
- What's the future vision of the e-evaluation programme should be taken on?

## IV. RESEARCH AIMS

- To design a model for an e-course on the internet on object-based programming for the level four students, Computer Science Department (CSD).
- To Plan an e-programme on the course of object based programming to assess the performance of CSD electronically via the web.
- To identify which method works well with the students' polls, e-evaluation method or the traditional one.
- To investigate the students' viewpoints on e-evaluation method applied in this research to e-evaluate the students.
- To put a form of future vision of the e- evaluation to be acquired.

## V. SIGNIFICANCE OF THE RESEARCH

- Putting forward a model for an e-course via the internet beneficial to working out more e-courses in higher education.
- Endeavoring to draw up a future form to an e-evaluation programme that can contribute to eliminate difficulties associated with the traditional evaluation method represented in handing and collecting exam papers and estimating them in order to make best use of the teacher and learner's time.
- Studying the feasibility of the idea, analyzing its application and employing it to draw some conclusions in an attempt to contribute to the improvement and development of the educational process and to make best use of computer and the internet technologies in opening up new horizons and prospects in this field.

## VI. RESEARCH METHODS

The researcher will use the experimental method as it cares to study factors and variables that affect the phenomena or problem and changes in some of its aspects with other stables to reach the causative relationship between these variables.

## VII. RESEARCH SAMPLE

The sample of the research consists of 36 students picked out randomly from the level four students of CSD, at Alrass Faculty of Science and Art. The sample is divided at random into three groups; two experimental groups and one control group as follows:

The first Experimental Group: consists of (12) male students who got their learning through e-course technique and have been e-evaluated.

The second Experimental Group: is comprising of (12) male students that received their learning via e-learning but evaluated by the traditional evaluation method, i.e., paper examination.

The control Group: includes (12) students of both sex who have learned according to the traditional education and have been evaluated via the traditional evaluation method, i.e. paper examination.

## VIII. RESEARCH OUTLINE

This research is restricted to the second unit in Object Oriented Programming (OOP) course, which is a Programme using Action Script (PAS) being taught to the students of 4th level CSD, faculty of Science and Arts, Alrass.

- Assumptions
- Using e-tests to evaluate the students' cognitive attainment is better than using the traditional evaluation method (i.e, paper tests) and there is a statistically significant distinction at a rate of 0.05 of the arithmetic mean between the degrees of the three groups (the two experimental groups and the control one) in favour of the first experimental group.
- There is a statistically significant distinction at a level of 0.05 of the arithmetic mean between the degrees of the three groups (the two experimental groups and the control one) in the time of carrying out the test in favour of the first experimental group, which adopted e-evaluation method.
- The students' attitudes are positive towards e-learning using e-course and e-evaluation method followed in assessing the attainment level of the sample students.

## IX. RESEARCH TERMINOLOGY

### A. The Internet

It's a technology linked with millions of computers connecting and allowing people together around the world to exchange information and ideas (proven E, 1999)

It's a multitude of computers communicating with one another; where millions of computers exchange information via the multiplex World Wide Web (Farhan N, Rafiq D., 1998)

Procedure

### B. Planning the e-course and uploading it on the web

The researcher planned an e-course for the subject of OOP planned for the students of level four at computer science department, Alrass Faculty of Science and Arts in accordance with the criteria specified by E-learning Unit at the University for Planning E-courses. The e-course has been uploaded on the website of the following site:

http://mansvu.mans.edu.eg/moodle//mod/resource/view.php?id=2936

### C. The course is designed according to the substantial procedures and steps followed in this respect as follows:

Pinpointing the learning topic: The second unit in OOP course is entitled as "Programme using Action Script" (PAS) designed for the students of fourth level at computer science department.

### D. Analyzing the Subject Matter

The content is compiled according to the nature of the subject of which the topic of the course is selected. The researcher turned to the content of the university book relevant to the course and a simulation of the programme has been prepared in the form of an educational booklet before planning it into an e-programme.

## X. IDENTIFYING THE EDUCATIONAL AIMS OF THE PROGRAMME

The educational aims of the programme fall into three categories:

- Cognitive aims: that is concerned with information and facts.
- Psychomotor aims: that tackles manual skills.
- Emotional aims: that deals with attitudes and values

## XI. WRITING THE SUBJECT MATTER OF THE PROGRAMME

The content of the course has been drafted including a home page that will be displayed when the course website is uploaded (via the University Website). The page includes a number of entries to different parts of the course. The subject matter has been turned to jury to give their opinion about it before embarking on the e-programming process.

## XII. E-PROGRAMMING

The researcher employs the following pogrammes and languages to prepare the course:

- Sound Forge – Front page - Action Script- Switch MAX – flash CS2.

Evaluating the course and Turning it to the concerned experts:

- After preparing the course, the researcher referred it to a group of experts from university professors specialized in computer sciences to give their opinions on the feasibility of the programme.

## XIII. PLANNING THE COGNITIVE ATTAINMENT TEST

The aim has been pinpointed, i.e., assessing the sample students' attainment of the concepts encompassed in the e-course which is prepared and edited on the internet through course teaching.

Specifications and Timetables have been prepared according to the compiled information which involves the topics that the test should include learning results which should be tested in accordance with the educational aims and the proportional significance of the subjects (proportional weight), [5]. Test specification table has been prescribed as follows:

Determining the number of expected in the test questions:

TABLE: SHOWS THE PROPORTIONAL WEIGHTS OF ATTAINMENT TEST ON SAP:

| NO | Content | Aims level | | | Total of content weights |
|---|---|---|---|---|---|
| | | application | understanding | cognition | |
| 1 | Writing action code and the types of data. | 2 | 12 | 6 | 20% |
| 2 | Orders of Movie Control. | 3 | 18 | 9 | 30% |
| 3 | Orders of mathematical function. | 3 | 18 | 9 | 30% |
| 4 | Working with properties. | 2 | 12 | 6 | 2% |
| | Total weights of the aims. | 30% | 60% | 10% | 100% |

Identifying the Proportional Significance and Aims (Table of Aims)

TABLE SHOWS THE NUMBER OF QUESTIONS IN THE ATTAINMENT TEST ON (PAS):

| NO | Content | Aims level | | | Total of content weights |
|---|---|---|---|---|---|
| | | application | understanding | cognition | |
| 1 | Writing action code and the types of data. | 0 | 1 | 1 | 2 |
| 2 | Orders of Movie Control. | 1 | 3 | 2 | 3 |
| 3 | Orders of mathematical function. | 1 | 3 | 2 | 3 |
| 4 | Working with properties. | 1 | 2 | 1 | 2 |
| | Total weights of the aims. | | | | 10 |

- Phrasing the questions of the test in such a way that makes these questions easily-understandable and accurately specified.
- Giving instructions in the beginning of the test that show the aim underlying it and define questions and how to answer them. A simulation has been carried out and the validity of the test has been checked through measuring the validity of its subject matter by turning it to a number of specialists from the teaching staff majored in the field of specialization in this respect.
- Ensuring the stability of the test that has been given to a pilot population composed of 10 individuals with a 20 day interval between the first and the second test. Using Richardson' equation, stability co-efficient is found to be (0.86), an evidence to the stability of the test.
- Time duration of the test has been worked out according to the following equation:
- Time duration of the test = (The time the fastest student took + the time the slowest students took) /2. After working out time duration, the validity and stability co-efficient, thus, validates the test which is composed of 10 questions carrying two marks each.

### XIV. DESIGNING THE TEST ON COMPUTER

To design the test in its final form the researcher followed the following steps:

- Eight models of the test have been designed on computer making a password for each and a username. Each student logs unto the home page of the test where he / she is asked to click a button which generates a number of tests at random, where he/ she takes one of them at random. At the end, the mark and the percentage are given in a window prepared for this purpose.
- As for the students who are given their tests according to the traditional evaluation method (paper tests), each was asked to pick a number randomly from a set of numbers and then he / she answer the random test. The test is corrected according to the correction key prepared for this purpose.
- Preparing a questionnaire based on the students' attitudes towards e-evaluation: This opinion poll aims at recognizing the viewpoints of the fourth level students sample consisting of 25 students of both sex at Computer Department, faculty of science and art as for their attitudes towards e-evaluation method. The questionnaire includes 20 phrases in connection with the test's range of accuracy its significance in assessing what it's built for, the effectiveness of the test's procedures, and its role in saving time and effort for each of the examiner and the examinee.
- The questionnaire has been referred to a group of experts in order to give their opinion with regard to its programme phrasing. The researcher worked out the percentage of the experts' approval of each phrase ruling out the points which didn't receive 85% at least of the jury's approval.

### XV. DISPLAYING RESEARCH RESULTS

*A. The 1st Assumption*

Using an e-test to evaluate the students' cognitive attainment is better than the traditional evaluation method (i.e paper test). There is a statistically significant distinction at a rate of 0.05 between the arithmetical mean of the degrees of the three groups in favour of the first experimental group.

- To test the validity of this assumption, the value of (X2) for the statistical significance between the arithmetical means of the two experimental groups and the control group has been worked out by using "kruskal wallis" co-efficient for three independent and identical samples of the cognitive attainment test degrees. The following table shows it as follows:

- The value of (X2) for the statistical significance of the arithmetical means of the degrees of the two experimental groups and the control one in the cognitive attainment test.

TABLE NO (3): SHOWS THE VALUE OF X2 FOR THE TWO EXPERIMENTAL GROUPS AND THE (CONTROL ONE) IN THE COGNITIVE TEST

| Group | Rank of means | (fd) | kruskal Wallis value (X2) | Sig. |
|---|---|---|---|---|
| The 1st experimental group | 29.54 | 2 | 19.45 | A statistical significance at a rate of 0.000 in favour of the 1st experimental group |
| The 2nd experimental group | 13.63 | | | |
| The Control group | 12.33 | | | |

Results indicate that there is a statistically significant distinction between the arithmetical mean of the students' degrees of the two experimental groups and the control one in favour of the first experimental group students' degrees. The distinction is essential and it doesn't trace back to the work of chance as kruskal Wallis' value of (X2) hit (19.945) to be bigger than the value of (X2) which struck (2.7) of rate of 0.000 and a freedom degree of 2.

It's noticed that mean of the attainment degrees of the second experimental group $\bar{x}$ equal ( 13.63) occurs after the first experimental group while the control group comes at last as for the mean of the test degrees $\bar{x}$ equal ( 12.33) and this is because the first and the second experimental groups received e-learning programme.

The preponderance of the first experimental group over the second experimental one in the test is due to the use of e-evaluation which helps students to remember and retrieve the information easily. Furthermore, the programme motivated them to improve their performance in the following question, the feeling that fosters examinee's self-confidence that helps him/her get better marks. The examine, moreover, transfers the experience he/she got while dealing with the e-course to the e-test he/ she takes finding no difference between what they learnt and what they have been evaluated in, the matter that the traditional evaluation method students didn't boast as they didn't have the opportunity to learn via a prepared e-course programme.

*B. The 2nd Assumption*

There is a statistically significant distinction at a level of 0.05 between the arithmetical mean of the degrees of the three groups(the two experimental groups and the control one) in the test's time duration in favour of the 1st experimental group which uses e-test in evaluation.

To test the validity of that assumption, the (x2) value of the distinction between the arithmetical means of the two experimental groups and the control one has been calculated by using kruskal Wallis co-efficient of their independent samples. The following table shows this as follows.

TABLE NO (3) (X2) VALUE (KRUSKAL WALLIS) OF THE DISTINCTION BETWEEN THE MEANS OF THE DEGREES OF THE TWO GROUP (THE TWO EXPERIMENTAL GROUPS AND THE CONTROL ON) ABOUT ATTAINMENT TEST'S TIME DURATION:

| Group | Rank of means | df | kruskal wallis value (X2) | Sig. |
|---|---|---|---|---|
| The 1st experimental group | 10.46 | 2 | 19.517 | A statistical significance at a rate of 0.000 in favour of the 1st experimental group |
| The 2nd experimental group | 16.08 | | | |
| Control group | 28.96 | | | |

N1= N2 =N3 = 12

Results indicate that there's a statistically significant distinction between the arithmetical mean of degrees of the two experimental groups and the control one in favour of the 1st experimental group, and that this distinction is not attributed to coincidence as the calculated kruskal's value of X2 hit (19.52) bigger than x2 value which reached (2.7) at a level of 0.000 and freedom degrees 2.

This points out that the first experimental group which used e-test in evaluating students' cognitive attainment, finished their test in time duration less than the other two groups. This is due to flexibility of dealing with the test and the clarity of its instructions. Moreover, providing the test with a direct feedback to the answers helps students move quickly from one question to the following and vice versa, the matter that results in concluding the test more quickly if compared to the traditional paper tests which fall short of achieving it.

*C. The 3rd Assumption*

"The students attitudes towards e-learning using e-courses and e-tests designed to assess the cognitive attainment level of the student sample are positive"

XVI. THE VALIDITY OF THE ASSUMPTION

To test the validity of this assumptions, the researcher, uses the general method of calculating X2 of the frequency table 1 ×2 to work out the statistical significance of frequency distinctions between the students approvals and disapprovals of each phrase of the questionnaire around e-evaluation method used in evaluating cognitive attainment. The following table shows this as follows:

TABLE NO (4) SHOWS FREQUENCIES, PERCENTAGE, X2 VALUE, AND THE STATISTICAL SIGNIFICANCE OF STUDENTS' RESPONSES TO THE FIRST PIVOT CONCERNED WITH "THE EFFECTIVENESS OF THE PROGRAMME USED IN EVALUATION".

| NO | Phrases | agree | | disagree | | X2 |
|---|---|---|---|---|---|---|
| | | f | % | F | % | |
| 1 | The e-test assess the students' cognitive attainment more accurately than the paper one | 19 | 76% | 6 | 24% | 6.76 |
| 2 | The test highly | 21 | 84% | 4 | 16% | 11.56 |

| NO | Phrases | f | % | F | % | X2 |
|---|---|---|---|---|---|---|
| | depends on accuracy | | | | | |
| 3 | The method of the test is more effective efficient and it's time and effort saving for both the examiner and the examinee than paper tests | 20 | 80% | 5 | 20% | 9 |
| 4 | The questions are couched in such a clear and simple way that's apprehensible to the learner | 18 | 72% | 7 | 28% | 4.84 |
| 5 | The test is comprehensive; covering all parts of the course. | 22 | 88% | 3 | 12% | 14.44 |
| 6 | I feel that the answer technique followed in the test is up-to-date. | 24 | 96% | 1 | 4% | 21.16 |
| 7 | The method of the test copes with e-methodology. | 22 | 88% | 3 | 12% | 14.44 |

The pervious table shows that x2 value of all phrases concerned with the students' view points on the first pivot of the questionnaire that deals with "the effectiveness of the programme used in evaluation " is bigger than the x2 value which struck 3.84 at a statistically significant rate of 0.05 and (1) freedom degree.

This indicates that the distinction between the observed frequencies and the expected ones around the phrases concerned with that pivot is statistically significant and it doesn't, therefore, go back to chance factor. The distinction between the students' approvals or disapprovals to the phrases of the first pivot in the questionnaire is in favour of the use of e-evaluation method.

On other words, the students agreed that e-evaluation method is effective, efficient and saving effort and time for the examinee and that the style of e-evaluation copes with the style of e-learning followed in the e-course. They also described the test as a developed one that doesn't allow cheating. In addition, there is a general satisfaction among the students due to the non-intervention of the human element in evaluation process. Thus, students feel completely satisfied with the result. They maintained that the test is in general better than paper test method used in the past. The student, furthermore, strongly stressed the importance of using e-evaluation method in all tests not only in the test of the student's performance marks of the year applied in this study.

TABLE NO (5) SHOWS THE FREQUENCIES, PERCENTAGES AND THE VALUE OF X2 AND ITS STATISTICAL SIGNIFICANCE OF THE STUDENTS' RESPONSE TO THE FIRST PIVOT CONCERNED WITH THE VALIDITY OF E-TESTING.

| NO | Phrases | Agree | | Disagree | | X2 |
|---|---|---|---|---|---|---|
| | | f | % | F | % | |
| 1 | The test offers the opportunities to try the wrong answers again. | 20 | 80% | 5 | 20% | 9 |
| 2 | I think that multiple correct answers fares well during the test. | 7 | 28% | 18 | 72% | 4.84 |
| 3 | I prefer the technique of putting out the result immediately. | 18 | 72% | 7 | 28% | 4.84 |
| 4 | The style of the test is developed and bars cheating. | 24 | 94% | 1 | 4% | 21.16 |
| 5 | I feel that the non-intervention of the human element in evaluation process is much better. | 18 | 72% | 7 | 28% | 4.84 |
| 6 | I think that the possibility of correcting the answers helped me improve my score | 18 | 72% | 7 | 28% | 4.84 |
| 7 | I feel stress and fear during carrying out my test. | 8 | 32% | 17 | 68% | 3.24 |
| 8 | Reinforcement is direct, exciting and untraditional. | 19 | 76% | 6 | 24% | 6.76 |
| 9 | I'm satisfied with getting my score without the intervention of the examiner | 23 | 92% | 2 | 8% | 17.64 |
| 10 | Using objective question in the test is much more better than open-ended questions followed in paper test | 21 | 84% | 4 | 16% | 11.56 |
| 11 | I think that e-evaluation method should be used in all tests not only in student performance score of the year. | 25 | 100% | - | 0% | 25 |

The previous table shows that the value of x2 of all the phrase concerned with the students' responses on the second dimension of the questionnaire that deals with "the validity of e-testing" is bigger than the value of x2 which struck 3.84 at a statistically significant rate of 0.05 and a freedom degree of 1. This indicates that the distinction between the observed frequencies and the expected ones on the phrases in respect is statistically significant and not attributed to coincidence factor.

The results indicated that the students agreed that the questions of the test had been phrased in a clear and simple way and that the test covers all parts of the course. The students also think that direct evaluation and giving them the opportunity to correct their answers helped them improve their scores, procedures that are not used in the traditional evaluation method (i.e, paper test).

Some students objected to giving more than one correct answer in the tests and complained that the instructions of the test didn't refer to this. Students also pointed out that direct evaluation of the questions made them feel excited, suspense and lack of traditionalism. The students, in addition, stated their un satisfaction with fill in the space questions because e-evaluating these questions may be inaccurate due to the difficulty of finding the typical answer. They also adopt the same attitude towards open-ended question in the questionnaire that deals with their viewpoints in general.

In connection to the open-ended question, the students commented in phrases which were repeated in a few sheets as follows:

- The test depends highly on accuracy and this is the best of its characteristics.
- The best thing in this test is putting out the result immediately. (Test score)
- The best thing in this test is that the test and the questions are picked out randomly, the matter that decreases cheating.
- The sole obvious shortcoming in this test is that it concentrated on to one question type.
- The method of the test is developed well
- There is more than one correct answer to some questions but it's is preferred to have just one answer for the question.
- The possibility of correcting question answer and trying it again is considered one good feature of that test(the possibility of moving from one question to another and vice versa)
- "Help" should be annexed to home page to define the properties of the test and the buttons
- The test should be limited to definite time duration.
- This method is progressive and up-to date and it doesn't depend on the traditional method of the previous tests.
- It's preferred to train students on the test and the endorsement of the score should be related to time.

## XVII. THE PROPOSED FUTURE VISION OF AN E-TEST

- Introduction of the test and presentation technique:
- Designing the introduction in such an attractive way that attracts the learner's attention and galvanises his motivation.
- The introduction includes an emulation model putting forth the way of answering the test and a window is made for user data.
- Moving to a following page which includes the examinee's data (username- password) linked to the student's database.
- Linking the test with Question Bank) database with a button that generates the number of test questions randomly from the database according to the number of questions the examinee will determine.
- Moving to the start page of the test where a random test is generated according to the number of questions set before.
- Planning a feedback to give the learner the opportunity to try again incorrect answers.

## XVIII. DATA OF FINAL TEST RESULTS

- After concluding question answers a button is activated to send the student's result to his/ her database where the programme puts out his / her score in the designed place.
- The test is administered according to general time duration not to specify time for each part of the test where the programme is immediately shut down after the time is over.
- It's available to the learner to print a report.

## XIX. CONCLUSION

In the light of applying research procedures to the current population, the following conclusion can be drawn out:

- Ensuring the availability of electronic learning for the students via the internet and by the educational banks help galvanize / give an impetus to students' motivation and enthusiasm for learning.
- The use of e-evaluation provides students with a sense of satisfaction and helps them greatly to get better score as it's characterized by accuracy, objectivity and its use of simulation technique which helps students easily retrieve information.
- The students have a high opinion on learning style using e-course via the internet and also towards e-evaluation method.
- It's strongly recommended to train students to use the internet and how to deal with e-courses and e-tests before using and applying this type of tests.
- The use of modern technology in education results in a significant development in the educational process as well as in students' thoughts and attitudes. In other words, the use of modern technology contributes to improving the educational product and enhancing quality of the learner.
- The provision of an electronic means to evaluate the students makes it easier for the staffs do their duties and give them plenty of time to unleash their creativity and achievement in their field of specialization.
- Planning e-tests, e-grading and writing down the marks get rid the staff and their assistants of the concomitant difficulties concerning handing over and collecting exam sheets followed in the traditional evaluation method (i.e, paper test).
- The use of e-test results in saving the time of both the examiner and the examinees, the matter that allows them make maximum use of their time in other business.

## XX. RECOMMENDATIONS

- Generalizing the use of e-evaluation method to high education courses via the internet.
- Disseminating the culture of e-learning and e-evaluation amongst the staff and their assistants through holding training sessions and periodicals.

- Holding training sessions with the staff and their assistants around how to prepare e-questions, e-tests and how to design a site for them on the web.

- Training undergraduate students on how to deal with the internet in order to be able to deal with e-learning and e-evaluation.

- Expansion in building educational banks for high education curricula and the encouragement of designing e-course on the web.

- Studying the feasibility of the idea, analyzing its application, and benefiting of the proposed e-design in this study to plan similar models that may enhance e-evaluation.

- Establishing a centre for e-evaluation that includes a number of specialists in assessment, evaluation and curricula in addition to computer and education technology teachers concerned with developing the ongoing evaluation process inside the university and with setting up and promoting e-question banks in order to give the student access to continuous training on that type of questions via the internet.